\documentclass{aa}  
\usepackage{graphicx}
\usepackage{txfonts}
\usepackage{booktabs}
\usepackage{multirow}
\usepackage[title]{appendix}
\usepackage{amsmath}
\usepackage[utf8]{inputenc}
\usepackage{tabularx} 
\usepackage{makecell} 
\usepackage{array}   
\usepackage{threeparttable}

\usepackage[colorlinks=true, citecolor=blue]{hyperref}
\hypersetup{
    citecolor  = blue,    
    linkcolor  = blue,   
    urlcolor   = blue,  
    pdfborder  = {0 0 0} 
}
\begin{document}
\title{Understanding mechanisms underlying solar cycle predictability with a general framework}

\titlerunning{Solar cycle predictability in dynamo models}

\author{Binghang Li 
	\inst{1,2}, 
	Jie Jiang 
	\inst{1,2}
	\and
	Yukun Luo\inst{1,2}}
\authorrunning{Li, B., et al.}
\institute{School of Space and Earth Sciences, Beihang University, Beijing, People's Republic of China\\
	\email{jiejiang@buaa.edu.cn}
	\and
	Key Laborotary of Space Environment Monitoring and Information Processing of MIIT, Beijing, People's Republic of China}

\date{Received February 08, 2026; accepted April 14, 2026}

  \abstract
   {The large-scale magnetic field observed at the solar surface is produced by the interior dynamo process. Whether this surface field also provides the dominant seed for the subsequent dynamo cycle, however, remains controversial, with important consequences for the predictive skill of solar dynamo models.}
   {We investigate the physical conditions under which this predictive skill of the surface field arises in dynamo models within a general framework.}
   {By applying Stokes’ theorem to the magnetic induction equation, we establish a direct physical link between the surface magnetic field and the subsequent dynamo process. The dominance of the surface induction integral in the net toroidal flux generation rate provides a quantitative criterion for assessing dynamo predictability, which we apply to five representative dynamo models.} 
   {This general framework shows that the surface magnetic field acquires predictive power when the surface poloidal field is efficiently coupled back into the dynamo loop through flux-transport processes (e.g., meridional circulation), a condition that can be satisfied in both Babcock–Leighton (BL)–type and $\alpha-\Omega$ mean-field dynamo models. The framework further identifies a new condition under which the surface magnetic field acquires predictive power: namely, that it represents the radial component of the interior poloidal field, as in the original BL-type dynamo scenario. In addition, the non-zero net toroidal flux across different dynamo models supports its use as a proxy linking the interior toroidal field to surface flux emergence.}
   {}

   \keywords{Solar cycle -- solar dynamo -- Sun: interior -- Sun: magnetic fields -- Sun: photosphere
               }

   \maketitle
   \nolinenumbers

\section{Introduction}
The amplitude of the approximately 11-year solar cycle varies over a wide range, from extended periods of very low activity, such as the Maunder minimum, to episodes of unusually high activity, such as the modern grand maximum in the 1960s \citep{Petrovay2020,Usoskin2023}. This variability significantly affects the space environment near Earth and numerous aspects of human activity \citep{Clark2006}, motivating efforts to predict solar cycle amplitudes. Since the solar cycle arises from a dynamo process operating within the Sun, prediction schemes based on solar dynamo models with data assimilation are particularly appealing. The first such attempts were made at the end of cycle 23, when two groups predicted cycle 24 using flux-transport dynamo models \citep{Dikpati2006a, Dikpati2006b, Choudhuri2007, Jiang2007}. These efforts subsequently prompted debate regarding the feasibility and robustness of solar cycle prediction based on dynamo model outputs \citep{Tobias2006,2007Bushby&Tobias}. Nevertheless, research along this line has continued. In addition to flux-transport dynamo models \citep{Jiang2013,Sanchez2014,Macario-Rojas2018}, $\alpha-\Omega$ mean-field dynamo models have also been applied to solar cycle prediction \citep{Kitiashvili2008, Jouve2011, Kitiashvili2020}. This body of work raises a key physical question: what are the necessary physical conditions for data assimilation of the surface magnetic field to result in predictive dynamo simulations?

The polar magnetic field constitutes a major component of the solar surface field, particularly during cycle minimum, and exhibits a strong correlation with the amplitude of the subsequent solar cycle. \cite{1978Schatten} first attempted to interpret the polar-field precursor within the framework of a Babcock–Leighton (BL) dynamo \citep{Babcock1961, Leighton1969}. In this class of models, the toroidal magnetic field is generated from a pre-existing poloidal field through shearing by differential rotation (the $\Omega$-effect). The resulting toroidal flux subsequently emerges to form sunspots and is therefore expected to be correlated with the poloidal field present at the preceding cycle minimum, which is predominantly concentrated in the polar regions. Based on this dynamo interpretation, a number of subsequent studies have attributed the polar-field precursor relationship to the BL mechanism, regarding it as observational support for the operation of the solar dynamo in a BL regime \citep{1978Schatten, 2013Munoz-Jaramillo}.

However, \cite{2011Charbonneau&Barlet} (hereafter CB11) suggested that such an interpretation of the polar-field precursor is not necessarily conclusive. They investigated three distinct stochastic forcing dynamo models and found that the polar field exhibits predictive skill in both a mean-field $\alpha-\Omega$ model with meridional circulation and a BL dynamo model, with nearly identical correlation coefficients in each case. Notably, the predictive skill of the surface polar field disappeared in the $\alpha-\Omega$ model when meridional circulation was omitted. Thus, they concluded that the surface poloidal field only shows predictive skill if it is coupled back into the dynamo loop. At least part of the polar field should be submerged and carried into the tachocline to contribute to the induction of the toroidal magnetic component that will give rise to sunspots of the subsequent cycle. The condition can be satisfied in both BL and mean-field dynamo models. 

Moreover, interpretations outlined above are based on dynamo models in which the dynamo operates primarily in the tachocline. In such dynamo models, both mean-field and BL types, the surface magnetic field is transported by meridional circulation into the tachocline. There, it feeds back into the dynamo loop by acting as the source to be amplified through the radial shear to generate the toroidal magnetic field \citep{2011Pipin,2020Charbonneau}. However, in recently developed distributed-shear BL dynamo models \citep{Zhang2022, Zhang2024, Jiang2025}, toroidal flux is generated in the bulk of the convection zone by latitudinal shear on the internal latitudinal poloidal field. Owing to the imposed condition of vanishing radial diffusion of the surface magnetic field \citep{Zhang2022,Luo2026}, the polar field cannot be submerged by the meridional flow, and the surface radial magnetic field does not directly participate in the internal dynamo process. Consequently, based on this dynamo model, a natural question arises: whether the assimilation of the surface magnetic field into the BL dynamo can retain predictive skill, and if so, what physical mechanism underlies it.

Beyond the unresolved questions, the methodologies commonly employed to assess predictability in dynamo models also warrant reevaluation. The conventional approach typically involves two steps \citep[e.g.,][]{2008Yeates, 2011Charbonneau&Barlet, 2021Kumar}: first, introducing stochastic forcing into a dynamo model and performing long-term simulations; second, statistically analyzing correlations between the polar field strength at cycle minimum and the amplitude of the subsequent cycle inferred from the model output. This approach has several limitations when applied to the study of dynamo predictability. First, it primarily quantifies the correlation between the polar field at cycle minimum and the following cycle amplitude, rather than directly addressing whether the observed surface magnetic field actively provides the source information for the subsequent dynamo process. Second, the method requires long-term simulations for each dynamo model, which are computationally demanding and model dependent. Third, the inferred predictive skill depends sensitively on the level of stochasticity and nonlinearity prescribed in the model \citep{2011Charbonneau&Barlet, 2021Kumar}. In addition, because there is no unified parameter that directly quantifies solar cycle strength in dynamo models, different proxies are adopted across studies \citep{2008Yeates, 2011Charbonneau&Barlet, Guo2021, 2024Ghosh}, making consistent comparisons between models operating in different regions of the convection zone difficult. These limitations highlight the need for a more general framework that can be applied uniformly across dynamo models while directly probing the capability of predictive skill.

The lack of a unified parameter to quantify solar cycle strength in dynamo models largely reflects the open question of how emerging flux forms from the toroidal field \citep{Fan2021}. Motivated by the observed hemispheric coherence of active-region orientations over a solar cycle \citep{Hale1919}, with only a small fraction violating Hale’s polarity law, \cite{2015Cameron&Schussler} (CS15) proposed that the hemispheric net toroidal flux, defined as the azimuthal average of the toroidal field, is the relevant quantity for sunspot formation. They further apply Stokes’ theorem to the induction equation. Thus the net toroidal flux can be calculated from a closed line integral bounding the cross-sectional area of each hemisphere. Using solar observations, CS15 concluded that the net toroidal magnetic flux is determined by the emerged surface flux and thus can be calculated from the observed surface magnetic field distribution. Especially, the dominant contribution to the net toroidal flux arises from the roughly dipolar polar field, providing a natural explanation for the polar-field precursor relationship. \cite{2024Finley} applied this approach to a 3D MHD dynamo simulation of a cool star, in which the net toroidal flux in each hemisphere is directly accessible, and verified the results of CS15 that the surface observables capture the toroidal flux generation inside the convection zone in their dynamo model. In contrast, \cite{2024Pipin} applied the same approach to their mean-field dynamo model \citep{Pipin2023} and found that the net toroidal flux can only partially be estimated from the surface magnetic field distribution. These differing results suggest that the extent to which surface magnetic fields encode information about the net toroidal flux may depend sensitively on the underlying dynamo regime.

Although the methodology proposed by CS15 does not identify the detailed physical origin or the location of the toroidal flux generation, it provides a quantitative measure of the net toroidal flux. Thus the application of Stokes' theorem to the induction equation provides a general framework for linking the surface observables to the internal dynamo process, and hence for assessing predictability of dynamo models. Within this framework, the net toroidal magnetic flux integrated over the entire convection zone can be adopted as a unified proxy for the amplitude of the solar surface activity. Moreover, the extent to which this quantity is predominantly determined by the surface line integral associated with the observed surface magnetic field distribution can serve as a criterion for evaluating the predictive skill of a given dynamo model. 

The paper is organized as follows. In Sect. \ref{sect:framework}, we present the general framework for analyzing solar cycle prediction. In Sect. \ref{sect:dynamo}, we introduce five dynamo models with different types. In Sect. \ref{sect:Results}, we apply the proposed framework to these dynamo models to assess their predictive skill and to elucidate the underlying physical mechanisms. Finally, Section \ref{sect:Conclusions} summarizes the main results.


\section{The general framework for understanding solar cycle prediction}
\label{sect:framework}

   \begin{figure}[!ht]
   \centering
   \includegraphics[width=0.8\linewidth]{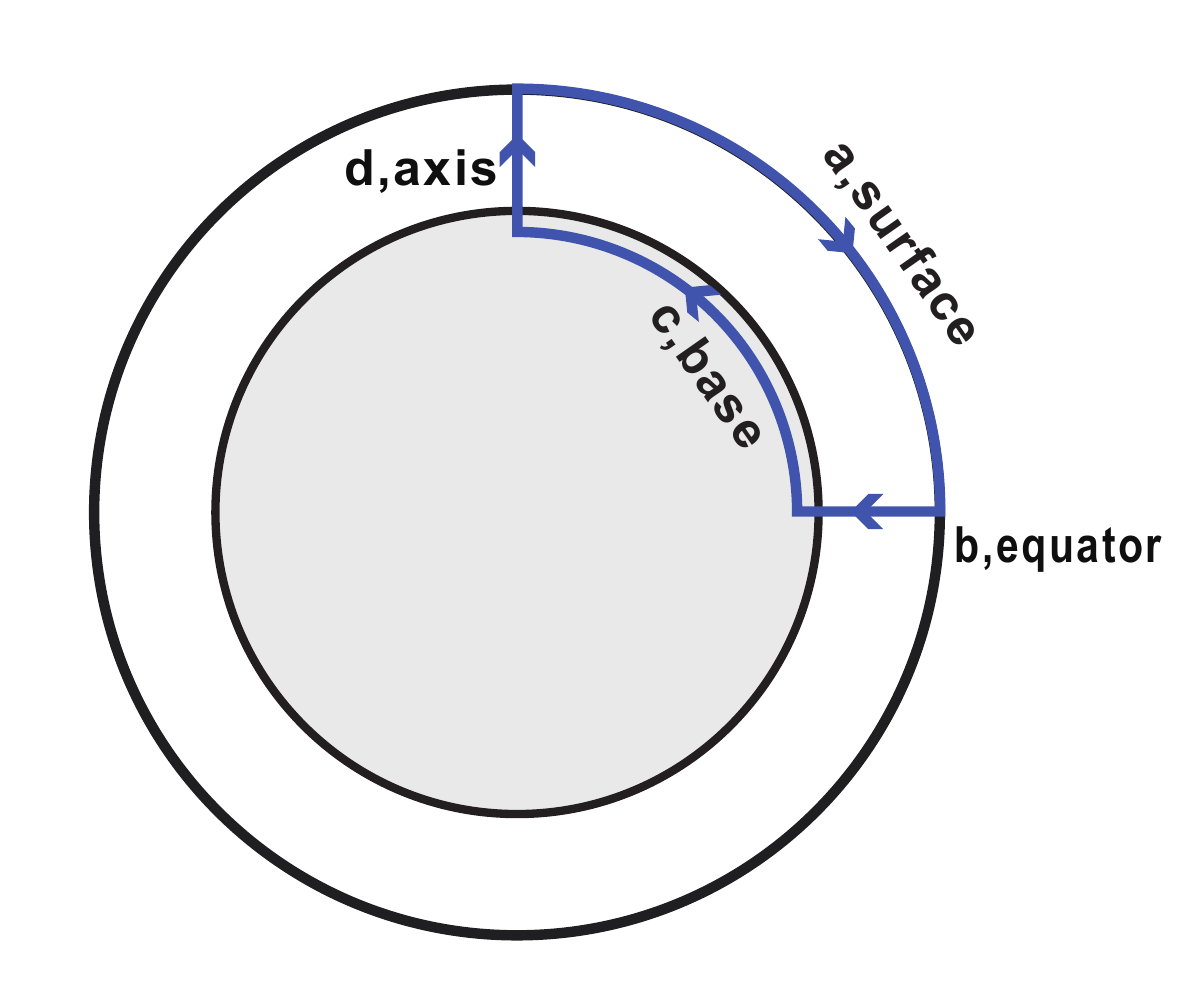}
   \caption{Schematic depiction of the application of Stokes' theorem. The white and gray zones represent the convection zone and the radiative zone, respectively. The blue contour $\mathcal{C}$ shows the path of integration around the area $\mathcal{S}$ in the northern hemisphere. And the contour $\mathcal{C}$ is clockwise and consists of the solar surface ($a$, surface), a radial segment in the equatorial plane ($b$, equator), a circular arc slightly below the bottom of the convection zone ($c$, base), a part along the axis of rotation ($d$, axis).}
  \label{fig:IntegContour}%
   \end{figure}
%


Since all dynamo models considered in the following sections are axisymmetric, we restrict our attention to the azimuthally averaged large-scale magnetic field. Its temporal evolution is governed by the mean-field induction equation \citep{1978Moffatt, 1980Krause&Raedler}, which can be written as follows:
\begin{equation}
\label{eq:induc}	
    \partial_{t}\overline{\boldsymbol{B}}=\boldsymbol{\nabla} \times\left(\overline{\boldsymbol{U}} \times\overline{\boldsymbol{B}}+\overline{\mathcal{E}}\right),
   \end{equation}
where $\overline{\boldsymbol{U}}$ and $\overline{\boldsymbol{B}}$ are the azimuthally averaged large-scale flow  field and magnetic field, respectively, and $\overline{\mathcal{E}} =\langle\boldsymbol{u}\times \boldsymbol{b}\rangle$ is the mean electromotive force (EMF) representing the azimuthally averaged correlation between the turbulent-scale fluctuating magnetic field $\boldsymbol{b}$ and velocity field $\boldsymbol{u}$. The term $\overline{\mathcal{E}}$ gives rise to turbulent magnetic diffusivity $\eta$, the $\alpha$-effect, and possible turbulent pumping. The $\alpha$-effect describes the regeneration of the poloidal field from the toroidal field and can arise from two distinct physical mechanisms, which are commonly used to distinguish different classes of dynamo models. In the $\alpha$-$\Omega$ mean-field model, the $\alpha$-effect originates from kinetic helicity in rotating, stratified convection. In contrast, in BL dynamos it results from the emergence, systematic tilt, and subsequent surface decay of sunspots.

We follow the approach of CS15 to calculate the net toroidal flux rate integrated in the whole convection zone by applying Stokes’ theorem to the induction equation (Eq. (\ref{eq:induc})). We define a contour $\mathcal{C}$ that encloses the area $\mathcal{S}$ in a meridional plane, as illustrated in Fig. \ref{fig:IntegContour}. The normal vector of the surface $\mathcal{S}$ is oriented parallel to the direction of the positive azimuthal field, $B_\phi$. The time derivative of the net toroidal magnetic flux, corresponding to the flux generation rate, in the northern hemisphere of the Sun, $\Phi_{tor}^N$, can be calculated as follows:
\begin{subequations}
   \label{eq:stokes}	
	\begin{align}
		\frac{\mathrm{d}\Phi_{tor}^N}{\mathrm{d}t} &= \frac{\mathrm{d}}{\mathrm{d}t} \int_{\mathcal{S}} B_{\phi} \mathrm{d}S, \\
		&=\oint_{\mathcal{C}}(\overline{\boldsymbol{U}}\times\overline{\boldsymbol{B}} + \overline{\mathcal{E}}) \cdot d\boldsymbol{l},
	\end{align}
\end{subequations}   
where $\mathrm{d}S$ is the surface element of $\mathcal{S}$, and $d\boldsymbol{l}$ is the line element along $\mathcal{C}$.
The net toroidal flux of the southern hemisphere of the Sun can be calculated using a similar approach.

Following CS15, we calculate the right-hand side of Eq. (\ref{eq:stokes}b) in a reference frame co-rotating with the solar equator at an angular rate $\Omega^{E}$. Note that $\overline{U}_\phi=0$ along the rotation axis,  $\overline{\boldsymbol{B}_p}=0$ along the base line segment since the poloidal magnetic field does not penetrate the low-diffusivity radiative zone. The line integrals in Eq. (\ref{eq:stokes}b) can be expanded into the following form: 
   \begin{equation}
   \label{eq:method}
    \begin{aligned}
     \frac{d \Phi_{tor}^N}{d t} = \underset{I_{1}}{\underbrace{\int_0^{\pi/2} (\overline{U}_\phi - \overline{U}_{0\phi})\overline{B}_r R_\odot d\theta|_{r=R_\odot}}} + \underset{I_{2}}{\underbrace{\int_{r_i}^{R_\odot} (\overline{U}_\phi - \overline{U}_{0\phi})\overline{B}_\theta dr |_{\theta = \frac{\pi}{2}}}}\\
     + \underset{I_{3}}{\underbrace{\int_{r_i}^{R_\odot} (\overline{\mathcal{E}}_r |_{\theta=0} - \overline{\mathcal{E}}_r |_{\theta=\frac{\pi}{2}})dr+ \int_0^{\pi/2} \overline{\mathcal{E}}_\theta R_\odot d\theta|_{r=R_\odot}-\int_0^{\pi/2} \overline{\mathcal{E}}_\theta r_{i} d\theta|_{r=r_{i}}}},
     \end{aligned}                
   \end{equation}
where $R_\odot$ denotes the solar radius, $\overline{U}_{0\phi}=R_\odot\sin\theta\Omega^{E}$ is the background rotation speed, and $r_{i}$ is the radial location of the base line segment $c$ in Fig. \ref{fig:IntegContour}. The value of $r_{i}$ is chosen as the first interior grid point above the bottom boundary and depends on the specific dynamo model considered in the following calculations. The integrals $I_{1}$ and $I_{2}$ represent the line integrals of $\overline{\boldsymbol{U}}\times\overline{\boldsymbol{B}}$ evaluated along the surface and equatorial segments of the integration contour, respectively. They quantify the magnetic induction produced by differential rotation acting on the large-scale magnetic field. Hereafter, $I_{1}$ and $I_{2}$ are referred to as the surface and equatorial induction integral, respectively.  In parallel, the integrals $I_{3}$ represent the contributions from the turbulent EMF. Specifically, when an isotropic scalarization parameter is employed, $\overline{\mathcal{E}}$ in Eq.\ref{eq:method}  
can be calculated by the following equation: 
\begin{subequations}
    \begin{align}
		\overline{\mathcal{E}}_r=\alpha \overline{B_{r}}- \eta \frac{1}{r \sin \theta} \frac{\partial}{\partial \theta} \left( \sin \theta \overline{B}_\phi \right), \\
	\overline{\mathcal{E}}_\theta=\alpha \overline{B_{\theta}} - \gamma\overline{B}_\phi + \eta \frac{1}{r} \frac{\partial}{\partial r} \left( r \overline{B}_\phi \right).
	\end{align}
\end{subequations}
It includes both turbulent diffusion, pumping effect, and field-generation processes associated with the $\alpha$-effect, which originates from kinetic helicity in $\alpha$-$\Omega$ mean-field dynamo models and sunspot emergence and evolution in BL dynamos. Note that pumping is assumed to be vertical, which means the density pumping is neglected \citep{Petrovay1994}. Hereafter, $I_{3}$ is referred to as the total EMF integrals.

In CS15, the primary objective was to estimate the net toroidal flux within a hemisphere using only observational data: the surface differential rotation and the magnetic field distribution. To achieve this, they introduced key approximations. Specifically, the equatorial induction integral $I_{2}$ along the equator was assumed to vanish, justified by helioseismology observations indicating that the angular velocity is nearly constant along the equator \citep{1998Schou}. Furthermore, the contribution of the total EMF integrals $I_{3}$ was simplified to consider only diffusive decay, represented as an exponential decay term with e-folding time $\tau$. Under these assumptions, $\frac{d \Phi_{tor}^N}{d t}$ is determined almost exclusively by $I_{1}$. Consequently, CS15 suggested that the surface induction integral $I_{1}$ captures the majority of the net toroidal flux generation, and since the polar field dominates $I_{1}$, this provides the physical basis for the polar-field precursor. 

However, CS15 inferred from observations that surface magnetic fields dominate the net toroidal flux generation, but this may not hold across all dynamo regimes. As noted by \cite{2024Pipin}, in some dynamo configurations the contributions $I_{2}$ and $I_{3}$ may reach magnitudes comparable to the surface induction integral $I_{1}$. The variability motivates us to apply the framework to assess predictability in different dynamo models. Dynamo simulations provide full access to the internal toroidal flux, allowing the net toroidal flux generation rate to be decomposed into the three contributions $I_{1}$, $I_{2}$, and $I_{3}$. Within this framework, the assessment of predictive skill can be formulated as a clear physical criterion: does the surface induction integral $I_{1}$ dominate the total generation rate of the net toroidal flux? We will apply this criterion to assess the predictive skill of the different dynamo models presented in the following section.

  
\section{Dynamo Models}  
\label{sect:dynamo}
In this section, we introduce the five dynamo models. The first three models employ parameter ranges similar to those of the models considered by CB11. The remaining two models are distributed-shear Babcock–Leighton (BL) dynamos recently developed by \citet{Zhang2022} and \citet{Jiang2025}; one includes meridional circulation, while the other does not.

\subsection{Model equations and boundary conditions}  
The axisymmetric large-scale magnetic field we investigate in Eq. (\ref{eq:induc}) from Sect. \ref{sect:framework} can be expressed in spherical coordinates as:
   \begin{equation}
     \overline{\boldsymbol{B}}(r, \theta, t)=B_\phi(r, \theta,t)\hat{\mathbf{e}}_{\phi} +\nabla\times\left[A(r, \theta,t)\hat{\mathbf{e}}_{\phi}\right],
   \end{equation}
where $B_\phi(r,\theta,t)\hat{\mathbf{e}}_{\phi}$ represents the toroidal field and $\nabla\times\left[A(r, \theta,t)\hat{\mathbf{e}}_{\phi}\right]=\boldsymbol{B}_p$ represents the poloidal field. For kinematic dynamo, the large-scale flow fields $\overline{\boldsymbol{U}}$ are assumed to be steady and can be prescribed as:
 \begin{equation}
     \overline{\boldsymbol{U}}(r, \theta)=r \sin \theta \Omega(r, \theta) \hat{\mathbf{e}}_{\phi}+\boldsymbol{U}_{p}(r, \theta),
   \end{equation}
where $\boldsymbol{U}_{p}(r, \theta)$ represents meridional circulation and possible radial pumping, and $\Omega(r, \theta)$ is the angular speed. The $\alpha-\Omega$ type dynamo equations, which describe the spatiotemporal evolution of the poloidal and toroidal fields, are derived by expanding the mean-field induction equation (Eq. (\ref{eq:induc})) in the azimuthal direction and the meridional plane, and can be written in the following form:
\begin{equation}
\label{eq:At}	
\frac{\partial A}{\partial t}+\frac{1}{s}\left(\boldsymbol{U}_{p} \cdot \nabla\right)(s A)=\eta\left(\nabla^{2}-\frac{1}{s^{2}}\right) A+S(r,\theta,B_\phi),\\
\end{equation}
\begin{equation}
\label{eq:Bphit}		
\begin{aligned}
    \frac{\partial B_\phi}{\partial t}+\frac{1}{r}\left[\frac{\partial\left(r U_{r} B_\phi\right)}{\partial r}+\frac{\partial\left(U_{\theta} B_\phi\right)}{\partial \theta}\right]=\\ \eta\left(\nabla^{2}-\frac{1}{s^{2}}\right) B_\phi+
    s\left(\boldsymbol{B}_{p} \cdot \nabla \Omega\right)+\frac{1}{r} \frac{d \eta}{d r} \frac{\partial(r B_\phi)}{\partial r},
    \end{aligned}
\end{equation}
where $s \equiv r \sin\theta$, $U_{\theta}$ represents the latitudinal component of the meridional circulation, while $U_{r}$ represents the radial component plus possible radial pumping $\gamma$, and $\eta$ is the turbulent diffusivity. The source term $S(r,\theta,B_\phi)=\alpha B_\phi$ describes the $\alpha$-effect in mean-field dynamo models, while it represents the regeneration of the poloidal field at the solar surface in BL dynamo models. All parameters including $\Omega(r, \theta)$, $\boldsymbol{U}_{p}(r, \theta)$, $\eta$, and the source term $S(r,\theta,B_\phi)$ for the five dynamo models presented in Sect. \ref{sec:DynamoModels} are summarized in the upper part of Table \ref{Table_1}.

The computational domain for mean-field models is $0.6R_{\odot} \le r \le R_{\odot}$ and $0 \le \theta \le \pi$, whereas BL models adopt a bottom boundary at $0.65R_{\odot}$. Regarding the boundary conditions, both classes of models share a perfect conductor condition at their respective lower boundaries, requiring $A = 0$ and $\partial (rB_\phi)/\partial r =0$; at the poles, $A = B_\phi = 0$ for both \citep{2020Charbonneau}. Two kinds of outer boundary conditions are considered. For the first three models described in Sect. \ref{sec:DynamoModels}, a potential-field condition with no electrical currents at $r=R_{\odot}$ \citep{1995Dikpati&Choudhuri} is imposed. For the remaining two models, we adopt a vertical-field condition, defined by $\partial (rA)/\partial r = 0$ and $B_\phi = 0$ at $r=R_{\odot}$, together with near-surface magnetic pumping. This setup of the outer boundary ensures that the resulting surface magnetic field evolution is consistent with surface flux transport models \citep{2012Cameron,2023Yeates}. 

The initial magnetic field is set with $\boldsymbol{B}_p=0$, retaining only a latitudinally dependent $\boldsymbol{B}_{\phi}$, and is then relaxed to cyclic solutions for each dynamo model presented in Sect. \ref{sec:DynamoModels}. The first three models are calculated using the code SURYA developed by A.R.Choudhuri and his colleagues \citep{Dikpati1994,Chatterjee2004}, while the last two are solved with the
Crank-Nicolson scheme combined with an approximate factorization technique \citep{Houwen2001} developed at Beihang University.

\subsection{Different dynamo models}
\label{sec:DynamoModels}

\subsubsection{Model 1: Flux transport BL model}
Flux transport dynamo (FTD) models were the earliest dynamo models used to explain polar-field precursor and are widely recognized for their predictive capability. The model developed by \cite{1999Dikpati&Charbonneau} is a representative example of this class. The parameters in the dynamo equation (Eqs. (\ref{eq:At}) and (\ref{eq:Bphit})), are adopted from the first case listed in their Table 1. Note that the original model yields symmetric solutions within one hemisphere, we extend the solution to a global domain ($0 \le \theta \le \pi$) to incorporate possible cross-hemispheric flux exchange by $I_{3}$, while maintaining this north-south symmetry. Model 2 and Model 3 are also extended to obtain global solutions for the same reason, compared with the original model. This setup is essentially equivalent to Model 3 in CB11 except for the computational domain.

In the model, the surface poloidal field is proportional to the toroidal field strength at the same latitude at the tachocline, with a classical algebraic $\alpha$-quenching as the nonlinear mechanism. The meridional circulation transports the surface poloidal flux toward the poles and subsequently downward to the tachocline, where it is stretched by differential rotation to generate the toroidal field for the subsequent cycle. Previous studies have demonstrated a correlation between the polar field at cycle minimum and the subsequent cycle strength by introducing stochastic fluctuations in the poloidal source terms \citep[e.g.,][]{2008Yeates, 2011Charbonneau&Barlet}. Applying the general framework described in Sect. \ref{sect:framework} to this dynamo model therefore provides a useful test of its ability to assess the predictive skill of a given dynamo configuration.

\subsubsection{Model 2: $\alpha-\Omega$ mean-field model with meridional circulation}
Our second model is an $\alpha-\Omega$ mean-field model that includes meridional circulation. The $\alpha$-effect is confined to the radial range [0.7 $R_\odot$, 0.8 $R_\odot$] and to a mid-latitude band ($\pi/4 \le \theta \le 3\pi/4$), with algebraic $\alpha$-quenching nonlinearity. The $\alpha$-effect is co-spatial with the region of toroidal field generation. All parameters froms are identical to those of Model 2 in CB11 to enable direct comparison, but the values are chosen to obtain a 11-year cycle resembling the real sun. Specifically, we take $\alpha_{0}$=0.25 m s$^{-1}$, $\eta_{0}=0.5\times 10^{11}$ cm$^2$ s$^{-1}$, $u_{0}=15$ m s$^{-1}$, which means $C_{\alpha}=\frac{\alpha_0R_\odot}{\eta_0}$=3.48, $R_{m}=\frac{u_0R_\odot}{\eta_0}$=209 instead of the value $C_{\alpha}$=0.5 and $R_{m}$=2500 adopted by CB11. The meridional circulation leads to the surface poloidal field to feedback into the region of the toroidal field generation. Hence in CB11, the model also shows the predictive skill. The presence of meridional circulation allows the surface poloidal field to feed back into the toroidal field generation region, and this model therefore exhibits predictive skill in CB11. Applying our framework enables us not only to quantify the contribution of the surface magnetic field to the net toroidal flux generation, but also to compare it directly with the contribution arising from the $\alpha$-effect.

\subsubsection{Model 3: Classical $\alpha-\Omega$ mean-field model}
Our third model is a classical $\alpha-\Omega$ mean-field dynamo operating without meridional circulation. We adopt the same parameters as Model 1 in CB11, except for a slightly decreased $\alpha_{0}$ amplitude. Specifically, we take $C_\alpha = 4.65$, compared to the value 5.0 used in CB11. In the absence of meridional circulation, there is no mechanism to feed the surface poloidal field back into the dynamo, and the model therefore does not exhibit a polar-field precursor, as demonstrated in CB11. This model allows us to investigate its behavior and contrast it with Model 2 within our general framework.

\subsubsection{Model 4: Distributed-shear BL model}
\label{sec:DynamoModel4}
Previous three dynamo models are consistent with the ones considered by CB11. Recently \cite{Zhang2022} and \cite{Jiang2025} developed a new type of BL dynamo, distributed-shear BL model. The model differs fundamentally from the FTD models in that the subsurface meridional circulation in such a model plays a minimal role. While the surface radial field originates from the evolution of sunspots at surface as the BL type, it does not exert a back-reaction process by meridional circulation due to the vertical outer boundary condition and near-surface pumping. A central objective of including this model is to determine whether the surface field in such a BL dynamo retains its predictive skill for the subsequent solar cycle. We adopt the solution of reference case in \cite{Zhang2022}.

\subsubsection{Model 5: Distributed-shear BL model without meridional circulation}
Meridional circulation in dynamo models not only couples the surface poloidal field back into the dynamo loop but also concentrate the surface magnetic flux toward the poles. This polar dominance has been central to previous interpretations of the polar-field precursor, since a polar-concentrated field substantially enhances the surface induction integral $I_{1}$ in Eq. (\ref{eq:method}). If the surface magnetic field distribution is no longer polar dominated, whether predictive skill is retained becomes a key question to address. 

In FTD models, the meridional circulation is a necessary ingredient. In contrast, distributed-shear BL dynamos can operate without meridional circulation, as it is not required for sustaining cyclic surface magnetic field evolution \citep{Leighton1964, Leighton1969, Baumann2004}. This distinction motivates the construction of Model 5, which is a distributed-shear BL dynamo without meridional circulation. The source term is identical to that used in Model 3, with parameters set to $\alpha_{0}$=0.2m s$^{-1}$, $\eta_{CZ}=3.2\times 10^{10}$ cm$^2$ s$^{-1}$, $\eta_{s}=1.15\times 10^{12}$ cm$^2$ s$^{-1}$. All other parameters are kept the same as in the reference case of \citet{Zhang2022}.

\begin{table*}[h]
\caption{Summary of Key Parameters, Governing Integration terms, and Predictive Skills of Selected Solar Dynamo Models.}
\centering
\begin{threeparttable} 
\renewcommand{\arraystretch}{1.2} 
\begin{tabularx}{\textwidth}{>{\raggedright\arraybackslash}X *{5}{>{\centering\arraybackslash}X}}
\toprule 
 & \textbf{Model 1} & \textbf{Model 2} & \textbf{Model 3} & \textbf{Model 4} & \textbf{Model 5} \\ \midrule 

\textbf{Parameters:} & & & & & \\ \addlinespace[1ex] 
$\begin{array}{@{}l@{}} 
    \Omega \\  U_p \\   \eta \\  S \\    \gamma  
\end{array}$ & 
$\begin{array}{@{}l@{}} 
    \text{Eq. (4) of DC99} \\
    \text{Eq. (5) of DC99} \\
    \text{Eq. (6) of DC99} \\
    \text{Eq. (7) of DC99} \\
    \multicolumn{1}{c}{/}
\end{array}$ 
 & $\begin{array}{@{}l@{}} 
    \text{Eq. (20-21) of PC20} \\
    \text{Eq. (5) of DC99} \\
    \text{Eq. (22) of PC20} \\
    \text{Eq. (5-8) of CB11} \\
    \multicolumn{1}{c}{/}
\end{array}$  & $\begin{array}{@{}l@{}} 
    \text{Eq. (20-21) of PC20} \\
    \multicolumn{1}{c}{/} \\
    \text{Eq. (22) of PC20} \\
    \text{Eq. (5-9) of CB11} \\
    \multicolumn{1}{c}{/}
\end{array}$&$\begin{array}{@{}l@{}} 
    \text{Eq. (10-11) of ZJ22} \\
    \text{Eq. (12) of ZJ22} \\
    \text{Eq. (14) of ZJ22} \\
    \text{Eq. (6-9) of ZJ22} \\
    \text{Eq. (13) of ZJ22}
\end{array}$ &$\begin{array}{@{}l@{}} 
    \text{Eq. (10-11) of ZJ22} \\
    \multicolumn{1}{c}{/} \\
    \text{Eq. (14) of ZJ22} \\
    \text{Eq. (6-9) of ZJ22} \\
    \text{Eq. (13) of ZJ22}
\end{array}$ \\ \addlinespace[3ex] 

Integration term dominating $\frac{d\Phi^N}{dt}$ & $I_{1}$ & $I_{1}$ & $I_{2}$ + $I_{3}$&$I_{1}$ &$I_{1}$ \\ \addlinespace

Has the predictive skill or not? & \text{Yes} & \text{Yes} & \text{No} & \text{Yes} & \text{Yes} \\ \addlinespace

Physical mechanism for the predictive skill & The surface poloidal field is efficiently coupled back into the dynamo loop & The surface poloidal field is efficiently coupled back into the dynamo loop & {/} & The surface magnetic field represents the radial component of the interior poloidal field & The surface magnetic field represents the radial component of the interior poloidal field \\ 
\bottomrule 
\end{tabularx}

\begin{tablenotes}[flushleft]
    \small
    \item \textbf{Note:} DC99: \cite{1999Dikpati&Charbonneau}; PC20: \cite{2020Charbonneau}; CB11: \cite{2011Charbonneau&Barlet}; ZJ22: \cite{Zhang2022}.
\end{tablenotes}

\end{threeparttable} 
\label{Table_1}
\end{table*}

\section{Results}
\label{sect:Results}

In this section, we apply the general framework presented in Sect. \ref{sect:framework} to the five kinematic axisymmetric dynamo models to assess their predictive skill. Using the output of each dynamo model, we directly calculate the net flux generation rate within the northern hemisphere $\frac{\mathrm{d}\Phi_{tor}^N}{\mathrm{d}t}$ based on Eq. (\ref{eq:stokes}a). We also directly calculate the surface and equatorial induction integral $I_{1}$ and $I_{2}$, and the total EMF integrals $I_{3}$ based on Eq. (\ref{eq:method}). By combining the evolution of the surface magnetic field distributions with the poloidal field configurations at solar minimum, we examine the relative magnitudes and phase relationships between these integrals and $\frac{\mathrm{d}\Phi_{tor}^N}{\mathrm{d}t}$. This analysis enables a physical interpretation of the results and helps elucidate the mechanisms underlying the models’ predictive skill. The results for the five dynamo models are summarized in the lower part of Table \ref{Table_1}.

\begin{figure*}
    \centering
    \includegraphics[width=0.9\linewidth]{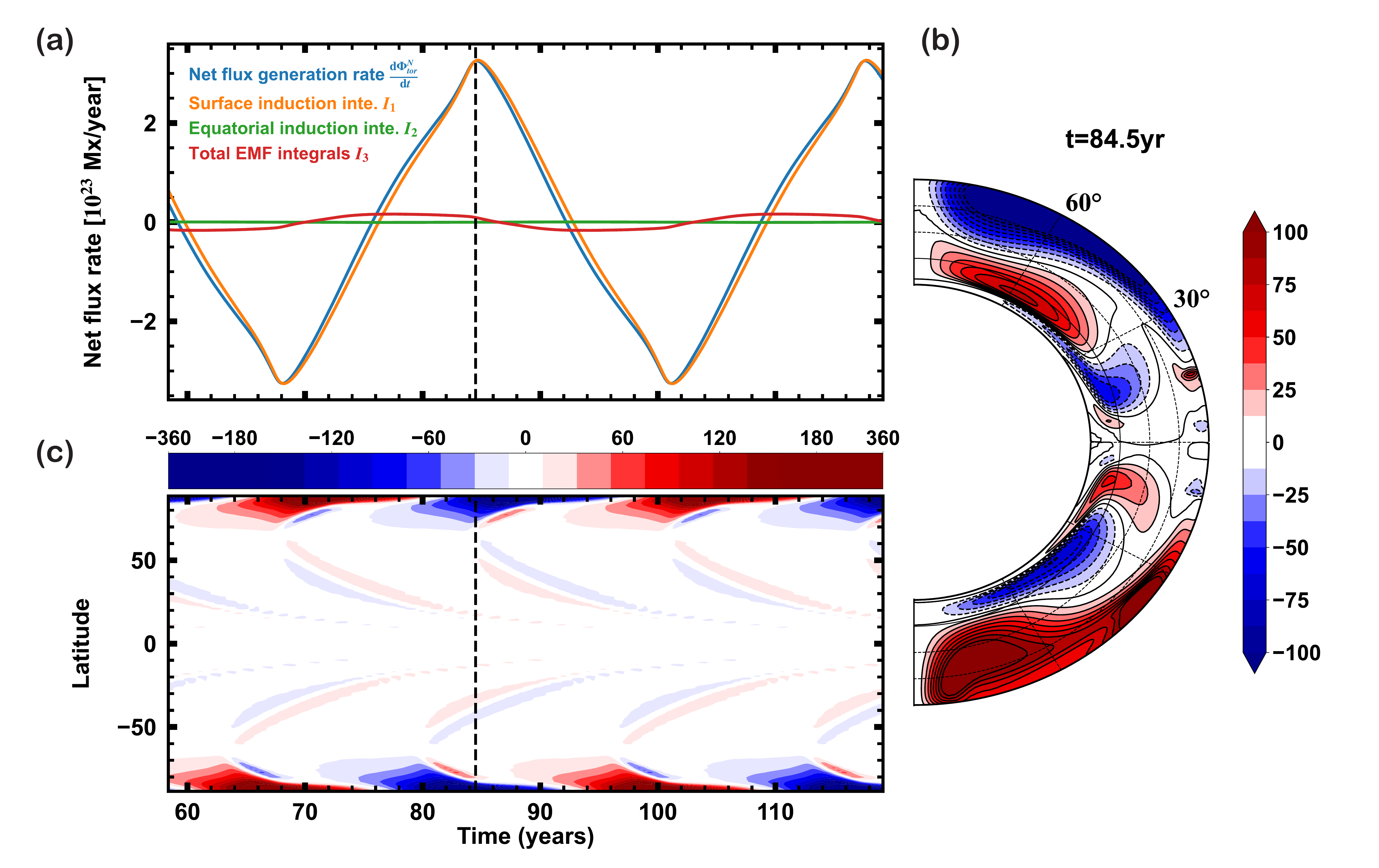}
    \caption{Results of Model 1, a flux transport BL dynamo model \citep{1999Dikpati&Charbonneau}. Panel (a): Temporal evolution of the surface induction integral $I_{1}$ (orange), equatorial induction integral $I_{2}$ (green), the total EMF integrals $I_{3}$ (red), and the net flux generation rate within the northern hemisphere $\frac{\mathrm{d}\Phi_{tor}^N}{\mathrm{d}t}$ (blue). Panel (b): Snapshot of the poloidal field configuration at cycle minimum. Panel (c): Time–latitude diagram of the surface radial magnetic field $B_r$. The vertical dashed lines in Panels (a) and (c) mark the timing of cycle minimum, as shown in Panel (b).} 
    \label{fig:model1}
\end{figure*}

\begin{figure*}
	\centering
	\includegraphics[width=0.9\linewidth]{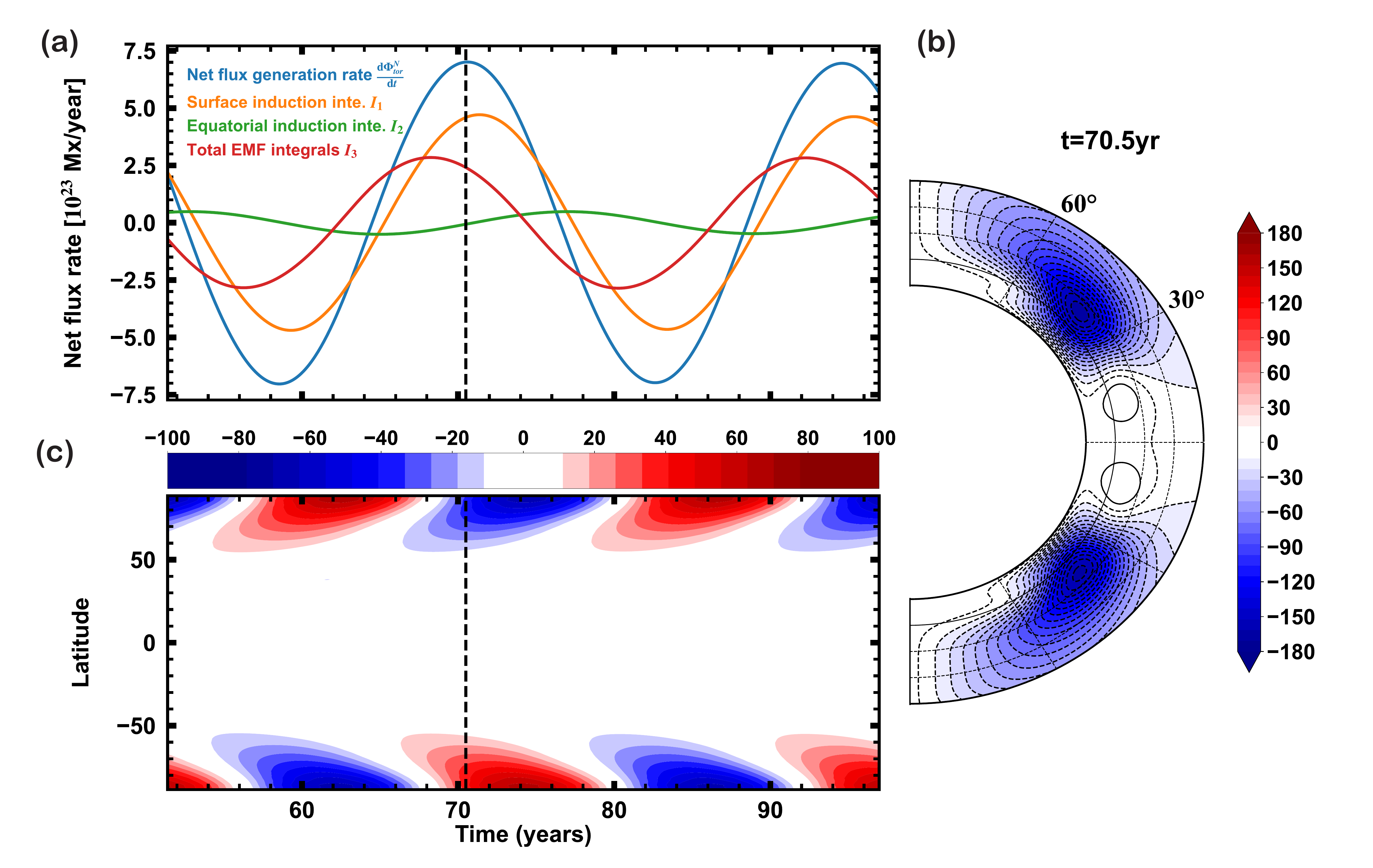}
	\caption{Same as Fig. \ref{fig:model1}, but for Model 2, an $\alpha-\Omega$ mean-field dynamo model with meridional circulation \citep{2011Charbonneau&Barlet}.}
	\label{fig:model2}
\end{figure*}

\begin{figure*}
	\centering
	\includegraphics[width=0.9\linewidth]{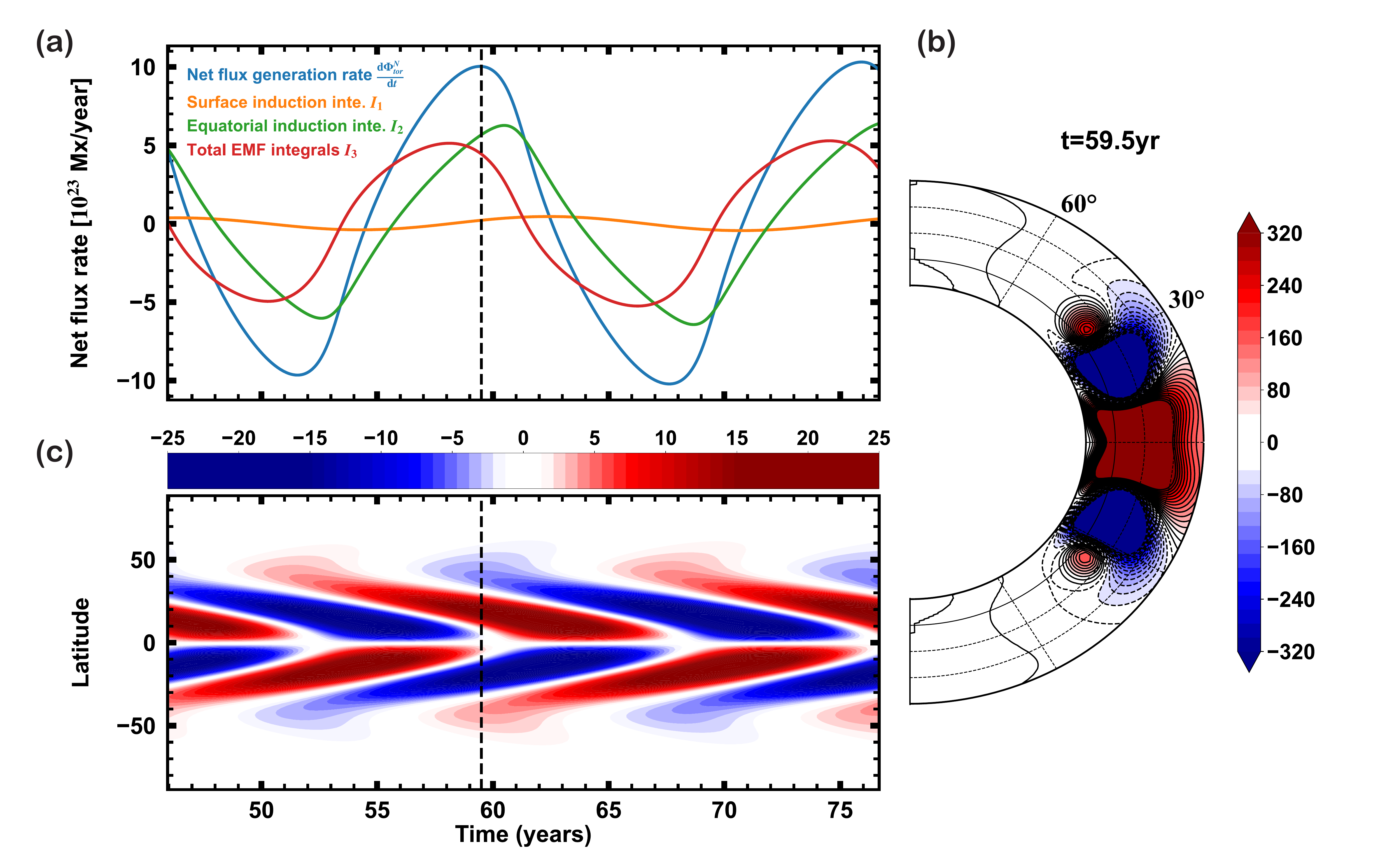}
	\caption{Same as Fig. \ref{fig:model1}, but for Model 3, a classical $\alpha-\Omega$ mean-field model \citep{2011Charbonneau&Barlet}.}
	\label{fig:model3}
\end{figure*}

Figure \ref{fig:model1} illustrates the results of Model 1, the flux transport BL model. Panel (a) shows the temporal evolution of the surface induction integral $I_{1}$, equatorial induction integral $I_{2}$, the total EMF integrals $I_{3}$, and the net flux generation rate within the northern hemisphere $\frac{\mathrm{d}\Phi_{tor}^N}{\mathrm{d}t}$ calculated based on the parameters and output of the dynamo model. As expected, the sum $I_{1}+I_{2}+I_{3}$ consistently reproduces $\frac{\mathrm{d}\Phi_{tor}^N}{\mathrm{d}t}$ in all models investigated below, validating our numerical implementation. For visual clarity and ease of comparison, we therefore display only $\frac{\mathrm{d}\Phi_{tor}^N}{\mathrm{d}t}$ in this and the following figures. Panel (a) shows that $I_{1}$ dominates $\frac{\mathrm{d}\Phi_{tor}^N}{\mathrm{d}t}$ and remains strongly phase-coherent with it throughout the time. Both $I_{1}$ and $\frac{\mathrm{d}\Phi_{tor}^N}{\mathrm{d}t}$ peak around solar minimum, reaching amplitudes of approximately $3.26 \times 10^{23}$ Mx/yr, while the contribution of $I_{2}$ and $I_{3}$ can be negligible. This demonstrates that, in Model~1, toroidal flux generation is overwhelmingly controlled by the surface induction term, explaining the strong predictive skill of this dynamo configuration.

This result can be understood from the configuration of the poloidal field at cycle minimum, shown in Panel (b), together with the surface magnetic field distribution in Panel (c). According to Eq. (\ref{eq:method}), the equatorial induction integral $I_{2}$ represents the flux generation captured by $B_{\theta}$ crossing the equator, while $I_{3}$ arises from the EMF. During solar minimum, there is essentially no $B_{\theta}$ crossing the equator, rendering $I_{2}$ negligible. Meanwhile, $\overline{\mathcal{E}}$ is dominated by the diffusion term, whose contribution is strongly suppressed by the efficient meridional circulation in this flux-transport dynamo model. Consequently, the contributions of both $I_{2}$ and $I_{3}$ to the toroidal flux generation are very small: $I_{3}$ reaches a maximum amplitude of only $1.63 \times 10^{22}$ Mx/yr and acts predominantly as a loss term, while $I_{2}$ remains close to zero. In contrast, the strong polar field evident in Panel~(c) dominates the surface poloidal field distribution, causing the surface induction integral $I_{1}$ to control the toroidal flux generation rate.

This outcome is consistent with a main result of CS15. $I_{1}$ effectively traces the peak in toroidal flux generation. This peak, which occurs around cycle minimum, largely determines the total net toroidal flux available for the subsequent cycle and thus governs the ensuing solar cycle. Through this physical linkage, the surface magnetic field, dominated by the polar field near solar minimum, acts as an effective precursor for predicting the strength of the next cycle, providing a physical explanation for the predictive skill of this dynamo model.

Figure \ref{fig:model2} illustrates the result of Model 2, an $\alpha-\Omega$ mean-field model with meridional circulation. As shown in Panel (a), the overall behavior is  similar to Model 1: $I_{1}$ ($4.59 \times 10^{23}$ Mx/yr) continues to dominate the sum, accounting for the majority of the net toroidal flux generation rate $\frac{\mathrm{d}\Phi_{tor}^N}{\mathrm{d}t}$ ($7.00 \times 10^{23}$ Mx/yr) and maintains phase coherence with it throughout the time. While $I_{2}$ remains near to 0, $I_{3}$ in this model is non‑negligible and provides a moderate positive contribution to the total toroidal flux generation, marking a departure from Model 1. 

This behavior can still be understood from the configuration of the poloidal field at cycle minimum (Panel b) and the surface magnetic field distribution (Panel c). As in Model 1, there is essentially no $B_{\theta}$ crossing the equator during solar minimum, and the differential rotation along the equator is weak, resulting in $I_{2}$ near to 0. However, in this $\alpha-\Omega$ mean-field model with meridional circulation, $\overline{\mathcal{E}}$ contains both inductive contributions from the $\alpha$-effect and turbulent diffusion. Unlike Model 1, where diffusion acts primarily as a loss term, the inductive component here provides a net positive contribution to toroidal flux generation. The presence of meridional circulation ensures that only a very small portion of the poloidal field across the equator and, more importantly, maintains the polar field’s dominance in the surface distribution. Consequently, $\frac{\mathrm{d}\Phi_{tor}^N}{\mathrm{d}t}$ is still primarily captured by $I_{1}$. This preserves the close correspondence between the evolution of the polar field and the amplitude of the subsequent cycle.

Crucially, these findings demonstrate that even when the poloidal field generation mechanism shifts from the BL mechanism to the internal $\alpha$-effect, the physical linkage established by applying Stokes' theorem to the induction equation remains valid, and the surface field dominated by polar field retains its predictive skill for the following solar cycle strength. This implies that the existence of polar-field precursor is not a definitive indicator of the BL mechanism's operation. The polar-field precursor phenomenon, in itself, shows no unique relation to the dynamo type and therefore cannot be used to conclusively constrain the dynamo type. This conclusion is in consistent with the statistical inferences from long-term simulations in CB11 and the analysis by \cite{2021Kumar}, both of which similarly caution against using the precursor correlation as a model discriminator.

Given that the poloidal field generation mechanism is not the decisive factor, we now seek to identify the fundamental physical driver of the predictive skill by the surface field. Classical $\alpha-\Omega$ mean-field models were widely recognized as lacking predictive skill \citep{2007Bushby&Tobias,Sanchez2014,2023Bhowmik}, which leads us subsequently investigate Model 3, a classical $\alpha-\Omega$ mean-field model, to verify its predictive skill within the general framework. In contrast to Models 1 and 2, the performance of Model 3 shown by Fig. \ref{fig:model3} reveals a fundamentally different regime. As shown by Panel (a), the net toroidal flux generation rate $\frac{\mathrm{d}\Phi_{tor}^N}{\mathrm{d}t}$ is found to be dominated not by $I_{1}$, but primarily by $I_{2}$ and $I_{3}$. Furthermore, $I_{1}$ loses its phase coherence with the flux generation rate, peaking instead around solar maximum. This shift is physically consistent with the poloidal field configuration at solar minimum (Panel b). In the absence of meridional circulation, the configuration possesses a significant $B_{\theta}$ crossing the equator and a strong poloidal field concentrated near the lower boundary below 0.7$R_\odot$. The former significantly enhances $I_{2}$, while the latter implies that $\overline{\mathcal{E}}$ is dominated by turbulent induction associated with the $\alpha$-effect. As a result, both $I_{2}$ and $I_{3}$ become substantial and can no longer be negligible. As these two terms exceed $I_{1}$, the evolution of the surface magnetic field becomes decoupled from the production of interior toroidal flux, explaining why the surface poloidal field in this model lacks predictive skill for the subsequent cycle \citep{Pipin2023}. 

Thus far, our analysis of the first three models obtain the consistent result with prior statistical approaches \citep{2011Charbonneau&Barlet}. This agreement both confirms that the polar-field precursor cannot be used to constrain the dynamo regime and demonstrates the effectiveness of the present framework. By replacing purely statistical correlations with a direct evaluation of the physical boundary integrals responsible for toroidal flux generation, the framework provides a physically grounded and robust measure of dynamo predictability.

In addition, Model 3 generates multiple bands of toroidal flux of alternating polarity within each hemisphere. It has been questioned whether the net toroidal flux, $\frac{\mathrm{d}\Phi_{tor}^N}{\mathrm{d}t}$, vanishes due to cancellation between these bands \citep{2024Finley}, and thus cannot serve as a proxy for surface flux emergence. However, we show that it retains a substantial value even in a classical $\alpha$–$\Omega$ mean-field dynamo. This result further supports the view that the net toroidal flux provides a physically meaningful link between the interior toroidal field and the emergence of magnetic flux at the solar surface.

\begin{figure*}
	\centering
	\includegraphics[width=0.9\linewidth]{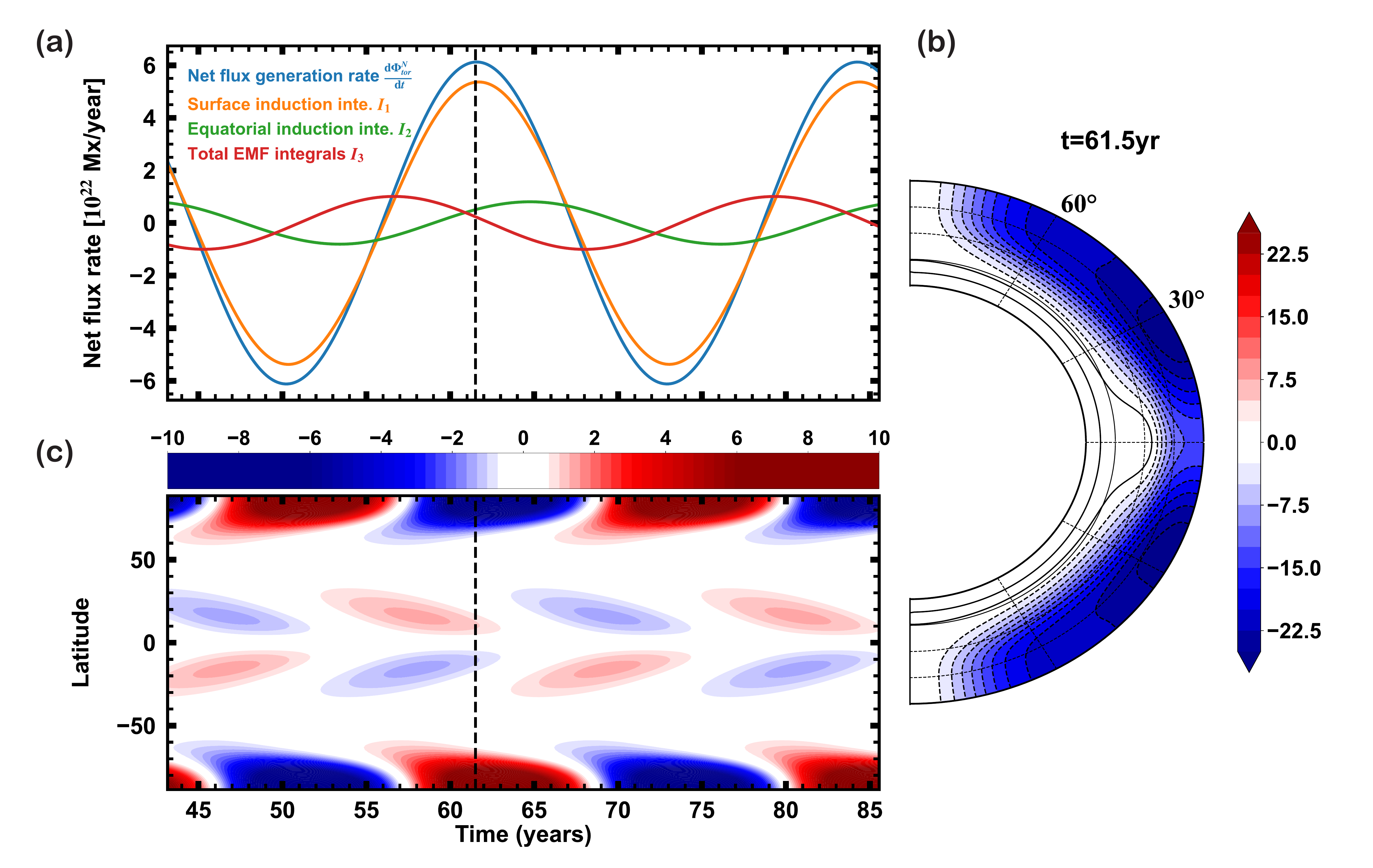}
	\caption{Same as Fig. \ref{fig:model1}, but for Model 4, a distributed-shear BL dynamo model \citep{Zhang2022}.}
	\label{fig:model4}
\end{figure*}

\begin{figure*}
	\centering
	\includegraphics[width=0.9\linewidth]{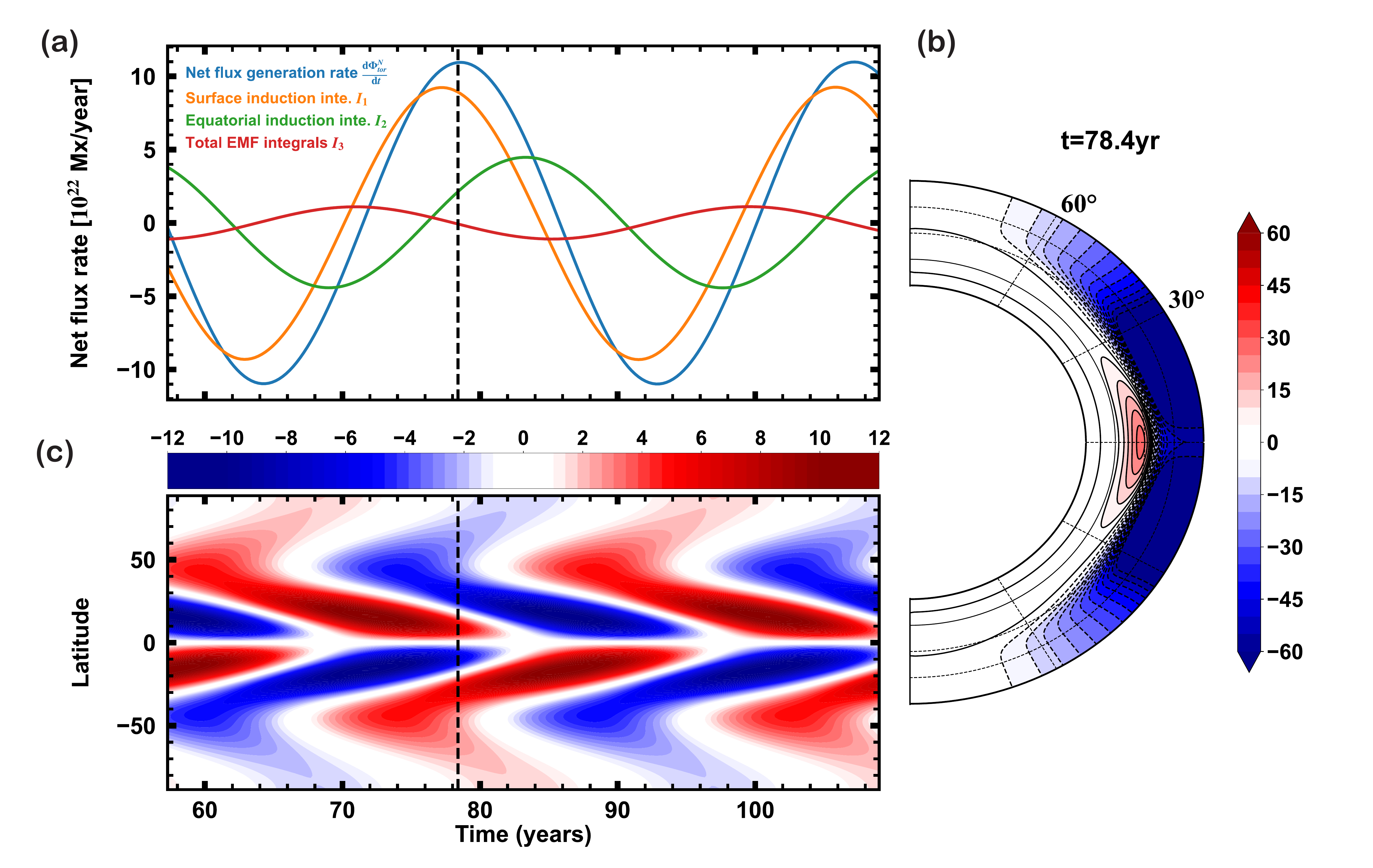}
	\caption{Same as Fig. \ref{fig:model1}, but for Model 5, a distributed-shear BL dynamo model without meridional circulation \citep{Zhang2022}.}
	\label{fig:model5}
\end{figure*}

The performance of the first three models highlights meridional circulation as a pivotal factor for the predictive skill of the relevant dynamo models. Traditionally, the primary role of meridional circulation is thought to be the transport of surface poloidal fields to the tachocline, where they are coupled back into the dynamo loop and act as physical source via radial shear \citep{2011Charbonneau&Barlet,2021Kumar}. In our framework, this process manifests as the suppression of $I_{2}$ and $I_{3}$ by minimizing cross-equatorial $B_{\theta}$. However, meridional circulation simultaneously reshapes the surface field into a polar-dominated configuration, thereby significantly enhancing $I_{1}$. The effect of the spatial distribution of surface field has not been isolated in previous studies. Furthermore, as discussed in Sect. \ref{sec:DynamoModel4}, emerging distributed-shear BL models rely on latitudinal shear throughout the convection zone to generate toroidal flux. In these regimes, even with meridional circulation, the surface field may not undergo a traditional feedback process due to the specific setup of the upper boundary. The predictive skill of such distributed-shear BL dynamos remains uninvestigated. To address these gaps, we employ Models 4 and 5, based on the distributed-shear BL framework developed by \cite{Zhang2022} and \cite{Jiang2025}, to isolate the role of surface field distribution and evaluate predictive skill in this dynamo regime. 

The behavior of Model 4 within the framework is illustrated in Fig. \ref{fig:model4}. Similar to Models 1 and 2, the surface induction integral $I_{1}$ accounts for the majority of the net toroidal flux generation rate $\frac{\mathrm{d}\Phi_{tor}^N}{\mathrm{d}t}$ ($5.36 \times 10^{22}$ Mx/yr out of $6.12 \times 10^{22}$ Mx/yr) and remains strongly phase-coherent with it throughout the cycle. In contrast to the earlier models, however, both $I_{2}$ and $I_{3}$ provide non-negligible contributions. These two terms operate largely in opposite phase and therefore partially cancel each other, such that $\frac{\mathrm{d}\Phi_{tor}^N}{\mathrm{d}t}$ remains dominated by $I_{1}$. This cancellation preserves the phase coherence between $I_{1}$ and $\frac{\mathrm{d}\Phi_{tor}^N}{\mathrm{d}t}$. Panel (b) shows that the poloidal field configuration at solar minimum presents a amount of cross-equatorial $B_{\theta}$, allowing $I_{2}$ to contribute positively to the net toroidal flux generation. At the same time, cross-hemispheric flux exchange acts as an efficient loss mechanism for toroidal flux \citep{Cameron2018,2024Finley}. And the turbulent EMF is dominated by the diffusive component, yielding a negative contribution from $I_{3}$.

Above results suggest that the predictive linkage between the surface field and the subsequent cycle amplitude remains effective in the distributed-shear BL dynamo model. Especially, the polar field also dominates the surface poloidal distribution in this regime as shown in Panel (c). Therefore, even in a model where the surface radial field does not directly serve as a source for the toroidal field and is not coupled back into the dynamo loop (due to the imposed vertical outer boundary condition), the surface field retains its role for evaluate the subsequent cycle’s strength.

Having established that the direct feedback of surface flux into the dynamo loop is not the decisive factor for the predictive skill of a dynamo model, we proceed to Model 5. By removing meridional circulation from Model 4, we aim to isolate and examine the specific influence of the surface magnetic field distribution. Unlike conventional tachocline-based models, the distributed-shear BL model explains the equatorward migration of toroidal flux through latitudinal dependence of the toroidal field generation \citep{Jiang2025}. Therefore, this model can regenerate the butterfly diagram without invoking the subsurface return flow of meridional circulation \citep{Babcock1961,Jiang2025}.

The behavior Model 5 is shown in Fig. \ref{fig:model5}. Compared with Model 4, the equatorial induction integral $I_{2}$ is significantly enhanced, while $I_{3}$ remains of comparable magnitude. This enhancement introduces a small phase offset between $I_{1}$ and the net toroidal flux generation rate $\frac{\mathrm{d}\Phi_{tor}^N}{\mathrm{d}t}$. Nevertheless, $I_{1}$ continues to account for the dominant fraction of $\frac{\mathrm{d}\Phi_{tor}^N}{\mathrm{d}t}$ ($9.26 \times 10^{22}$ Mx/yr out of $1.10 \times 10^{23}$ Mx/yr) and remains largely phase-coherent with it. As shown Panel (b), the absence of meridional circulation allows the cross-equatorial $B_{\theta}$ to accumulate, leading to the increase in $I_{2}$. In contrast, the cross-hemispheric flux exchange and the associated diffusive loss captured by $I_{3}$ remain comparable to those in Model~4. Since $I_{2}$ and $I_{3}$ contribute with opposite signs, their net effect does not significantly alter the dominant contribution of $I_{1}$.This result demonstrates that predictive skill does not require the polar field to be the dominant component of the surface magnetic field, as illustrated in Panel~(c). 

A contrasting behavior emerges when comparing Models 3 and 5. Despite exhibiting nearly identical surface magnetic field distributions characterized by mid-latitude maxima and dominant equatorward branches, Model~3 shows no predictive skill, whereas Model~5 retains it. The difference arises from the spatial relationship between the surface field and the dominant induction region. In the classical $\alpha-\Omega$ mean-field Model 3, the absence of meridional circulation leads to a strong accumulation of poloidal field near the lower boundary, which substantially enhances $I_{3}$ through positive $\alpha$-effect induction and reduces the relative contribution of $I_{1}$. As a result, the surface poloidal field becomes spatially decoupled from the deep radial shear layer where toroidal flux is generated, preventing the surface field from serving as an effective precursor. In contrast, Model 5 operates in the bulk of the convection zone, and its surface magnetic field represents the radial component of the interior poloidal field. 
The surface field is therefore intrinsically linked to the shear process occurring immediately beneath it, allowing $I_{1}$ to capture the net toroidal flux generation rate. We thus conclude that, in addition to the flux-transport feedback operating in Models 1 and 2 as proposed by CB11, a second physical mechanism can also ensure the predictive skill of a dynamo. Specifically, when the surface field directly reflects the interior poloidal field, it remains physically coupled to the dynamo process and serves as a reliable precursor.

\section{Discussion and Conclusions}
\label{sect:Conclusions}

In this study, we apply Stokes’ theorem to the magnetic induction equation, and establish a direct physical link between the surface magnetic field and the subsequent dynamo process. This direct connection builds a general framework that provides a quantitative criterion for assessing dynamo predictability by examining the magnitude and phase relationship between the surface induction integral $I_{1}$ and the net toroidal flux generation rate $\frac{\mathrm{d}\Phi_{tor}^N}{\mathrm{d}t}$. Grounded in robust physical principles, the framework is then applied to investigate five representative dynamo models, helping to elucidate the mechanisms underlying solar cycle predictability.

The net azimuthally averaged toroidal flux generation rate is adopted as the proxy for the solar surface magnetic field in the paper. This is motivated by observational arguments, as proposed by \cite{2015Cameron&Schussler}. This assumption has been questioned in the context of classical $\alpha$–$\Omega$ mean-field dynamo models, which often produce multiple bands of toroidal flux of alternating polarity within each hemisphere. \cite{2024Finley} suggested that there would be zero net flux in each hemisphere for such models. However, in our Model 3 (a classical $\alpha$-$\Omega$ mean-field dynamo), we still obtain a substantial net toroidal flux, even larger than in the other models considered in the paper. In addition, the existence of multiple toroidal-flux bands is not unique to classical $\alpha$-$\Omega$ mean-field dynamo. Flux-transport B-L dynamos, such as Model 1 and Fig. 11 of \cite{Dikpati2006b} also present similar multi-band structures. The presence of a non-zero net toroidal flux across different types of dynamo models suggests that it provides a physically meaningful link between the interior toroidal field and the emergence of magnetic flux at the solar surface.

Using the hemispheric net toroidal flux integrated over the entire convection zone as a proxy for the surface magnetic field does not account for the uncertain location of toroidal field generation (e.g., confined to the lower layers or distributed throughout), which remains an open question. The flux-emergence process is also not yet fully understood. Recent BL dynamo models \citep[e.g.,][]{Cameron2020,Cloutier2025,Zhang2026} have begun to incorporate simplified treatments of flux emergence, enabling a more direct link between the net toroidal flux and the surface field.

Our results show that predictive skill in the surface field, such as the polar-field precursor, is not a definitive indicator of the B-L mechanism and does not provide a conclusive constraint on the dynamo type. In addition, these results highlight that, for tachocline-based models to exhibit predictive skill, meridional circulation is crucial in coupling the surface magnetic field back into the dynamo loop. Model 3 (classical $\alpha-\Omega$ mean-field model) exhibits a regime lacking predictive skill, where the net toroidal flux generation rate is dominated by the contributions from $I_{2}$ and $I_{3}$. This is attributed to $B_{\theta}$ crossing the equator and strong poloidal field concentrated near the lower boundary below 0.7$R_\odot$, which significantly enhances $I_{2}$ and $I_{3}$, respectively. However, upon the introduction of meridional circulation in Model 2 ($\alpha-\Omega$ mean-field model with meridional circulation), the flow suppresses this bottom concentration and reduces the cross-equatorial $B_{\theta}$. Consequently, $I_{2}$ is significantly diminished and $I_{3}$ is partially weakened, thereby endowing the model with predictive skill. It is anticipated that as the meridional flow speed is further increased towards a fully advection-dominated regime, $I_{2}$ and $I_{3}$ would be further suppressed, eventually converging to the behavior of Model 1 (FTD). This physical transition aligns with the findings of \cite{Sanchez2014}, who find that the predictive skill is shown to decrease with the strength of $\alpha$-effect, and to increase with the strength of the imposed meridional circulation. This may also illustrate the nearly identical correlation coefficients between FTD and mean-field models with strong meridional circulation investigated by \cite{2011Charbonneau&Barlet}. 

Our results reveal a novel predictive mechanism distinct from the traditional process of physically coupling the surface magnetic field back into the dynamo loop. Our results show when the surface magnetic field represents the radial component of the interior poloidal field, the predictive skill also arises. In the distributed-shear dynamo model \citep{Zhang2022}, meridional circulation does not directly transport the surface magnetic field back into the internal dynamo loop. Remarkably, both models including Model 5, which operates without any meridional circulation, exhibit predictive skills. This comparative setup effectively rules out the necessity of a polar-dominated surface field distribution, a geometric factor often overlooked in previous studies. The underlying reason lies in the nature of these dynamos: the poloidal field manifests as a large-scale dipole structure, and toroidal flux generation relies primarily on latitudinal shear. In this configuration, $I_{1}$ capture $\frac{\mathrm{d}\Phi_{tor}^N}{\mathrm{d}t}$ by this natural physical correlation, endowing the models with inherent predictive skills. 

Contrary to previous assumptions \citep{2015Cameron&Schussler,2024Finley}, our results reveal that moderate cross-hemispheric flux exchange actually facilitates the estimation of $\frac{\mathrm{d}\Phi_{tor}^N}{\mathrm{d}t}$ by $I_{1}$. Specifically, when a cross-equatorial $B_{\theta}$ is captured by $I_{2}$, it simultaneously interacts with the latitudinal shear to generate toroidal fields of opposite polarity across the hemispheres. This also creates a strong diffusive gradient, which is captured by $I_{3}$ as a negative contribution. Consequently, $I_{2}$ and $I_{3}$ tend to  operate largely in opposite phase and therefore partially cancel each other. This cancellation preserves the phase coherence between $I_{1}$ and $\frac{\mathrm{d}\Phi_{tor}^N}{\mathrm{d}t}$. Given that the solar magnetic field at cycle minimum is dominated by large-scale antisymmetric axial components \citep{Luo2024}, such cross-hemispheric flux exchange likely plays a critical role in the high predictive skill observed in reality.

\begin{acknowledgements}
    We thank the anonymous referee for valuable comments and suggestions that have improved the manuscript. The research is supported by the National Natural Science Foundation of China (grant Nos. 12425305, 12350004, and 12173005) and China's Space Origins Exploration Program.
\end{acknowledgements}


\bibliographystyle{aa}
\bibliography{references}
\end{document}